\documentclass[letterpaper, 10 pt, conference]{ieeeconf}  

\IEEEoverridecommandlockouts                              
\usepackage[utf8]{inputenc}
\usepackage[T1]{fontenc}
\usepackage{graphicx}

\title{\LARGE \bf
Extended Reality (XR): The Next Frontier in Education
}

\author{Shadeeb Hossain$^{1}$ 
\thanks{*Research Division,Shadeeb Engineering Lab,Brooklyn,NY 11223}
\thanks{$^{1}$S.Hossain is the Principal Engineer and Founder of
        Shadeeb Engineering Lab, United States of America
        {\tt\small shadeeb@shadeebengineeringlab.com}}%
}

\begin{document}

\maketitle
\thispagestyle{empty}
\pagestyle{empty}

\begin{abstract}

Extended Reality (XR), encompassing Virtual Reality (VR), Augmented Reality (AR), and Mixed Reality (MR), is revolutionizing education by creating immersive, interactive learning environments. This article explores the potential of XR to enhance student engagement, experiential learning, and skill development while addressing the challenges of widespread adoption. High implementation costs, technical complexities, and ethical concerns especially regarding student privacy and biometric data protection still possess significant barriers to integration. The discussion also highlights regulatory compliance with GDPR and FERPA and the importance of cybersecurity frameworks to safeguard sensitive learner data. Ultimately, the article provides insights into balancing innovation with accessibility and ethics in the evolution of XR based education. 

\end{abstract}
\def\IEEEkeywords{\textit{Index Terms}---}

\vspace{2pt}
\noindent\IEEEkeywords Augmented Reality classroom, Extended Reality, SMART classrooms, Virtual Reality, XR-based learning

\section{FROM E-LEARNING TO XR ENABLED LEARNING }

\begin{figure}[thpb]
      \centering
      \includegraphics[width=\linewidth]{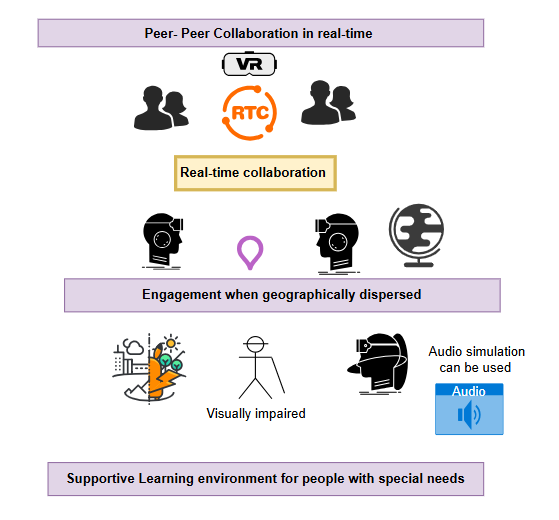} 
      \caption{:Several advantages of Extended Reality in educational settings: (i) Peer-Peer collaboration in real-time, (ii) engagement even if the learners are geographically dispersed, (iii) provide learning support to special needs individuals.}
      \label{fig:schematic}
\end{figure}

In the era of evolving SMART technology -enabled classrooms, with the adoption of (i) e-learning, (ii) adaptive learning and assessment practices, and (iii) differentiated learning approaches; it is not surprising that Extended Reality (XR) is emerging as the next step to facilitate student-centered learning\cite{c1,c2,c3}. Extended Reality (XR) is an umbrella term that integrates Virtual Reality (VR), Augmented Reality (AR), and Mixed Reality (MR). It encompasses real and virtual combined environments where human users interact seamlessly with computer generated inputs, enabling immersive and interactive learning experiences. 
XR offers several advantages in educational settings. First, it enables collaboration through shared virtual learning spaces where students and instructors can interact in real time. Second, XR fosters meaningful engagement between facilitators and peers, even when participants are geographically dispersed, thereby reducing barriers to access. Third, XR can provide tailored support for students with special needs by creating adaptive, personalized environments that accommodate diverse learning requirements. Finally, XR delivers cost-effective simulated environments for learning and training, offering students the opportunity to practice skills in safe, controlled scenarios that would otherwise be resource-intensive or difficult to replicate in traditional classrooms. Fig.\ref{fig:schematic} shows the schematic highlighting the several advantages of Extended Reality in educational settings. 

Especially in educational settings, collaboration across the learning community, such as communication between students, and facilitators, as well as peer-peer integration is integral to a comprehensive learning experience. Traditional classrooms allow for rich exchanges, but in digital and remote formats, such interactions are often limited to text-based chats, discussion boards, or video conferencing, which can lack spontaneity and depth of in-person engagement. Extended Reality (XR) is uniquely positioned to bridge this gap by creating immersive environments that simulate physical presence, foster collaboration, and restore the interpersonal dynamics of traditional classrooms within a digital framework. 

Augmented Reality (AR) and Artificial Intelligence (AI) are increasingly being recognized as powerful tools to support personalized and adaptive learning, particularly for individuals with diverse or special educational needs. By embedding AI into AR/ XR platforms, education can overcome key classroom challenges, such as providing real-time feedback, adaptive pacing, and individualized content delivery. For instance, AI-driven translation services within AR can assist non-native speakers by enabling real-time language support, thereby reducing barriers to comprehension and participation. Beyond translation, AI can also analyze learner interaction in AR environments to recommend tailored instructional strategies, helping students engage with complex concepts at their own pace. As AR and AI technologies continue to converge, they hold the potential to transform traditional learning spaces into highly inclusive, dynamic, and student-centered ecosystems. 

\section{EVOLVING AR HAPTIC SENSORY SYSTEMS: TOWARDS INCLUSIVE LEARNING ENVIRONMENT}


Virtual Reality (VR) can also support visually impaired learners by enabling kinesthetic learning experiences in a safe and controlled environment. For example, Maidenbaum et al. (2016) demonstrated the use of visual-to-audio Sensory-Substitution Devices (SSD) to create an alternative virtual learning environment accessible to blind users\cite{c3}. In their study, participants with cognitive blindness successfully navigated a virtual environment, located doors, distinguished them based on structural features and surrounding context, and then walked through them. The tasks achieved a 97\% success rate, highlighting the potential of VR to provide more challenging training opportunities through simulated environments and to expand accessibility for visually impaired learners. 

In addition to sensory substitution techniques such as SSD, more recent innovations include the Camera Input-Output (CamIO) system, which provides audio-based AR guidance for blind users\cite{c5}. CamIO directs users through simple verbal prompts to locate points of interest on the physical object. In one study with four blind participants, the system was evaluated by comparing the time required to identify target hotspots. Results showed that participants guided by CamIO performed 8.27 times faster than those without guidance. These findings highlight the potential of AR-based approaches such as CamIO in informing the design of next generation haptic sensory system for visually impaired learners.  

Beyond audio-based simulation prompts, researchers at the Keck School of Medicine, University of Southern California, demonstrated that adapted augmented reality (AR) glasses could enhance both mobility and gasping performance for individuals with low vision\cite{c6,c7}. Their results showed a 50\% improvement in mobility and a 70\% improvement in grasping accuracy. The system operates using simultaneous localization and mapping (SLAM), which generates real-time 3D reconstructions of the surrounding environment. These are then translated into colored visual overlays that highlight potential obstacles. Such approaches can be extended to support kinesthetic learners and further enhanced with Artificial Intelligence (AI) -driven prompts, thereby broadening accessibility for visually challenged learners. These trajectory points towards the design of multimodal haptic simulation systems tailored for diverse learning styles-for instance, visual learners (through AR eyewear) and auditory learners (through audio prompts, as discussed in the previous example).

\section{LEVERAGING GAMING OPTION TO IMPROVE LEARNING EXPERIENCE
}

Gamification refers to the application of game design principles and mechanisms to enhance the learning experience by making it more engaging, interactive and rewarding. In educational settings, gamification has shown promise in motivating otherwise disengaged learners by transforming traditional lessons into interactive, challenge-based modules that stimulate curiosity and participation. When combined with Extended Reality (XR) technologies, gamification further immerses students in interactive learning experiences, offering rewards, progress tracking, and problem-solving scenarios that mimic real-world challenges. 

Several pilot studies have demonstrated the potential of gamified XR training in improving learner engagement, knowledge retention, and skill acquisition, particularly in domains requiring repetitive practice or complex decision-making. For instance, a study by Zikas et al. (2016) examined the role of mixed reality (MR) games in improving the learning experience \cite{c8}. Conducted in a primary school history class, the study implemented Mixed Reality Serious Games and Gamification strategies that incorporated gesture-based learning, Meta AR-glasses, and integrated game shells to create a highly immersive and engaging learning environment. 
Similarly, three case studies by Stott and Neustadter (2013) explored the effectiveness of gamification and identified key factors that can provide guidelines for improving learning practices\cite{c9}. The studies highlighted that there is no single, universal model of gamification that guarantees classroom effectiveness, however, gamification consistently fosters a sense of ownership among learners, enhances motivation, and transforms static lessons into dynamic, interactive experiences. 

\begin{figure}[thpb]
      \centering
      \includegraphics[width=\linewidth]{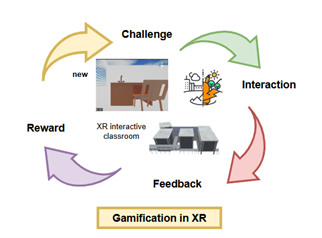} 
      \caption{Proposed Gamification steps for XR based learning system}
      \label{fig:schematic1}
\end{figure}

By integrating gamification with XR technologies, educators can develop highly personalized and motivating learning environments. These environments not only support diverse learner needs but also provide scalable frameworks for innovative teaching practices in both traditional and remote classrooms. Fig. \ref{fig:schematic1} shows a schematic of the different steps in a proposed XR based gamification system that includes : (i) introduction of a challenge, (ii) the learners interacting with the challenge environment, (iii) getting feedback on their performance, (iv) assigned rewards based on performance, and (v) finally getting new set of challenges.

\section{ADOPTION OF XR TECHNOLOGY IN K-12 AND HIGHER EDUCATION CLASSROOM LEARNING}
With the rise of XR and gamification technology, both K-12 institutions and universities have begun to integrate immersive learning tools into their classrooms. These technologies are transforming traditional lessons into interactive, experience-driven environments that promote deeper understanding and engagement. 
For instance, a pilot study by Hmoud et al. (2023) involving fourteen high school students examined the impact of XR technology on student engagement in Biology \cite{c10}. The study assessed four key dimensions of engagement - (i) cognitive, (ii) emotional, (iii) social, and (iv) behavioral to determine how immersive technologies influence learning outcomes. The results indicated a significant increase in cognitive engagement, suggesting that XR can help students better visualize and comprehend complex biological concepts through interactive simulations and 3D representations. 

Similarly, Batra et al. (2025) at Purdue University introduced XRXL, an extended reality system designed to enhance student engagement in large lecture environments \cite{c11}. This project considered one of the largest deployments of co-located collaborative XR applications, was evaluated through an IRB- approved mock lecture involving 82 student participants. The session focused on explaining the concept of neural networks through XR-driven visualization and interaction. Findings from the study demonstrated that XRXL not only increased student participation and attention but also improved conceptual understanding and collaborative engagement in large-scale instructional settings. 

These studies illustrate the growing acceptance of XR and gamification as transformative tools in both secondary and higher education, paving the way for immersive, personalized, and student-centered learning experiences. 

\section{CHALLENGES IN MAINSTREAM ADOPTION OF XR TECHNOLOGY IN EDUCATION SECTOR} 
There are several challenges associated with the widespread adoption of Extended Reality (XR) in the education sector. One of the primary challenges is the high cost of hardware and infrastructure required for immersive learning environment. XR based classrooms depend on a variety of specialized devices, including virtual reality (VR) headsets, customized haptic vests, simulators, and other accessories tailored to a specific lesson. Advanced XR headsets can cost approximately \$5000 each, while VR vests that provide sensory feedback can cost around \$900 per unit. Beyond hardware additional expenses arise from programming the VR environment, cloud computing, and storage, and cybersecurity protection to ensure data integrity and privacy. These initial costs can be further compounded by maintenance expenses, as frequent use by students often leads to equipment wear or accidental damage. 

Another significant challenge involves the training and professional development required for educators, administrators and support staff to effectively implement and supervise XR-based instructions. Many educators already face demanding workloads that include lesson planning, grading, classroom management, and administrative duties. The introduction of XR technologies add another layer of complexity, [potentially leading to training fatigue among faculty. Moreover, in a field already grappling with teacher shortages and high turnover rates, the need for specialized XR expertise may further complicate recruitment and retention efforts. Institutions may need to establish new qualification criteria or certification pathways for XR integration specialists, instructional technologists, and educators with immersive learning expertise.

\section{ETHICS AND PRIVACY CONCERNS IN XR}
In our previous work, we emphasized the critical importance of privacy and data protection within Extended Reality (XR) environments. Extended Reality (XR) systems inherently collect large volumes of biometric and behavioral data- including eye-tracking metrics, gesture and motion patterns, voice inputs, and even emotional or cognitive response indicators- to enable adaptive and personalized learning experiences. While these capabilities enhance interactivity and engagement, they also introduce significant privacy and ethical risks. 

To safeguard students’ rights, the collection and processing of such sensitive information must adhere to established data protection frameworks such as General Data Protection Regulation (GDPR) and Family Educational Rights and Privacy Act (FERPA). These frameworks govern how educational institutions handle personal and biometric data, ensuring that data collection is transparent, purpose-specific, and consent-driven. 

Without clear governance policies and robust cybersecurity infrastructure, student biometric data remains vulnerable to breaches, unauthorized access, or misuse for malicious purposes such as identity theft or profiling. Institutions adopting XR technologies must therefore implement strong encryption protocols, secure cloud storage practices, and continuous monitoring systems to mitigate cyber risks. Additionally, ethical considerations must extend beyond compliance-educators and developers should promote digital literacy and informed consent, helping students understand what data are being collected and how they are used. 
Ultimately, the responsible integration of XR into education requires a privacy-by-design approach, where ethical safeguards are embedded into every stage of system development and deployment. By prioritizing trust, transparency, and accountability, XR can evolve into a powerful yet responsible tool for immersive learning. 

\section{CONCLUSION: XR AND FUTURE OF IMMERSIVE LEARNING}

Extended Reality (XR) has emerged as a powerful tool capable of transforming education through immersive, interactive, and student-centered learning experiences. By integrating Virtual Reality (VR), Augmented Reality (AR), and Mixed Reality (MR), XR enables learners to visualize abstract concepts, engage in simulated practice, and collaborate in shared digital spaces beyond the boundaries of the traditional classroom.  The incorporation of gamification and Artificial Intelligence (AI) further amplifies XR’s potential -creating adaptive engaging, and personalized learning environments that respond dynamically to each student’s needs and learning style. 

However, the successful deployment of XR in educational systems requires addressing significant challenges related to cost, training, accessibility, and ethics. Institutions must invest not only in technology itself but also in the pedagogical frameworks and educator readiness that make XR integration meaningful and sustainable. Attention to privacy, inclusivity, and equitable access will be essential to ensure that XR technologies do not reinforce existing disparities but instead expand learning opportunities for all. 

The future of XR in education lies in the convergence of AI, cloud computing and 6G connectivity which has the potential to make immersive learning scalable, affordable and data driven. As these technologies evolve, classrooms of the future may transcend physical boundaries and create a connected learning ecosystem that fosters collaboration and experiential learning experiences.



\end{document}